\documentstyle[12pt]{article}

\begin{document}

\newcommand{\be}{\begin{equation}}
\newcommand{\ee}{\end{equation}}
\newcommand{\ba}{\begin{eqnarray}}
\newcommand{\ea}{\end{eqnarray}}
\newcommand{\f}{\frac}
\newcommand{\s}{\sqrt}

%\vspace*{4cm}
\begin{center}
{\bf FROISSART-MARTIN BOUND }

\vspace*{2Mm}
{\bf IN SPACES WITH COMPACT EXTRA DIMENSIONS}\\ 
\vskip \baselineskip
{\bf V.A.Petrov}\\

\vspace*{1mm}
{\it Division of Theoretical Physics, IHEP, 
Protvino, Russia}\\
\end{center}

\vspace*{5mm}
\begin{abstract}
We generalize the Froissart-Martin upper bound for the scattering
amplitude in
spaces with compactified extra dimensions. 
\end{abstract}

\vspace*{10mm}
Geometrically the Froissart-Martin upper bound in the Minkowsky
space-time of
arbitrary dimension $D$~[1,2] 
\be
\sigma^D_{tot}(s)\leq\, const\,(R_0(s))^{D-2} 
\ee
defines the maximal ``transverse area'' at which colliding particles
effectively feel each other. 

The question arises what happens with Eq.(1) if some of $D-1$ 
spatial\footnote{We don't consider time-like extra dimensions
(moreover
compactified), as in this case microcausality, a key ingredient for
analyticity,
become ambiguous if not worse.} dimensions are compactified? 

We find more transparent from the physics point of view to use impact
parameter
representation for the scattering amplitude, 
\be
T(s,\vec k_T)\simeq 4s
\int d^{D-2}Be^{i\vec k_T\vec B}\tilde T(s,\vec B). 
\ee
where $\s s$ is, as usual, the c.m.s. collision energy, and
$k_{T}^2\simeq -t$
is
related to the transferred momentum.

From the unitarity condition 
$$
Im\tilde T(s,\vec B)\geq \left |\tilde T(s,\vec B)\right |^2, 
$$
analyticity in the $t$-plane with a nearest singularity at $t=t_0>0$,
and
polynomial boundedness , $\left |T\right |<s^N$, one easily recovers
the bound~
(1), with $R_0(s)\simeq \f{N\log s}{\s{t_0}}$. 

Let us compactify for simplicity one transverse dimension in such a
way that
with $\vec B\equiv (\vec b, b_{D-2}\equiv \beta)$ one has 
$$
\tilde T(s,\vec b,\beta+2\pi R)=\tilde T(s,\vec b,\beta), 
$$
where $R$ is the compactification scale. 
Conjugated integration in $k_{D-2}$ converts into the sum 
$$
\sum^{\infty}_{n=-\infty}\f{e^{in\beta/R}}{2\pi R}. 
$$
$(D-2)$th component plays now the r\^ole of the source of an
additional mass
spectrum (twice 
degenerated under $n\to -n$) adding to the usual mass$^2$ the term
$n^2/R^2$.
If $R$
is
small (as it seems to be the case in reality) then these modifications
to the
spectrum do not influence the nearest singularity, and, hence,
analyticity. 

Below we consider the amplitude for scattering of ``light'' particles
with
$n=0$, which presumably correspond to usual observed particles. In
this case we
have from usual arguments
\ba
&&ImT(s,0)\leq 4s 
\int d^{D-3}b
\int\limits^{\pi R}_{-\pi R} d\beta\Theta
(R^2_0-b^2-\beta^2)=\nonumber \\
&&= 4s\f{\Gamma ((D-3)/2)}{2\pi ^{D-3}/2}
\cdot 2\pi R_0^{D-2}(s)\Phi (R_0, R,D), 
\ea
where 
\ba
\Phi & = &\int\limits^{1}_{0}
d\xi\xi^{D-4}(1-\xi^2)^{1/2}\Theta
\left (\xi  -\s{\max [
0, (1-\pi^2R^2/R^2_0(s)) ]}\right ) + 
\nonumber \\
& + & 
\f{\pi R}{R_0(s)}\cdot \f{(1-\pi^2R^2/R^2_0(s))_{+}^{D^{-3}}}{D-3}.
\nonumber 
\ea
At $R<< R_0(s)$ we get 
\be
Im T(s,0)\leq const \cdot s\cdot R^{D-3}_0 (s)\cdot R. 
\ee
In the opposite, somewhat academic, limit $R/R_0\to \infty$ we come
back to
Eq.(1)
$$
Im T(s,0)\leq const \cdot  s\cdot R^{D-2}_0(s). 
$$

It is evident that above derivation can be readily generalized to more
than one
compact extra dimensions. As well as that all the inference can be
made more
rigorous mathematically. 

I express my gratitude to A.V.Kisselev, A.Martin, A.A.Logunov,
N.E.Tyurin and
Yu.S.Vernov for discussions.

\enddocument